\RequirePackage{fix-cm} 
\documentclass[twocolumn,epjc3]{svjour3}
\usepackage[dvipsnames]{xcolor}
\usepackage{textcomp}
\usepackage{gensymb}
\usepackage{array}
\usepackage{amsmath}
\usepackage{amssymb}
\usepackage{graphicx}
\usepackage{microtype}
\usepackage[colorlinks=true, allcolors=blue]{hyperref}
\usepackage{xurl}
\usepackage{doi}
\usepackage{siunitx}
\usepackage{multirow}
\usepackage{booktabs}
\usepackage{makecell}
\usepackage{orcidlink}
\usepackage{flushend}
\usepackage{lineno}
\journalname{Eur. Phys. J. C}
\begin{document}
%
\title{Study of the \mbox{Run-3} muon flux at the SND@LHC experiment}
\author{The SND@LHC Collaboration\thanksref{e1}$^\text{\normalfont,1}$}
\thankstext{e1}{e-mail: \href{mailto:alexandra.serban@cern.ch}{alexandra.serban@cern.ch}}
\institute{Full author list at the end of the article.\label{1}}
\date{Received: date / Revised version: date}
%
%
\maketitle

\abstract{
\begin{sloppypar}
    Long-range muons produced in proton--proton collisions at the ATLAS interaction point constitute the primary background for neutrino interaction searches at the SND@LHC experiment. This work presents a comprehensive characterization of the muon flux throughout LHC \mbox{Run-3}, benchmarking Monte Carlo simulations against experimental measurements. Measured and simulated muon rates agree within 10--15\% across all \mbox{Run-3} configurations. Following the substantial background increase in 2024 as a result of a beam optics change, the reversion to nominal optics in 2025 did not restore the 2022--2023 levels due to the unprecedented adoption of horizontal crossing in ATLAS. As enlightened by simulation results, the latter enhanced the contribution of high-angle muons originating from diffractive proton losses in the LHC Dispersion Suppressor region. Their identification enabled the design of mitigation strategies that were experimentally validated. The simulation framework was also applied to the future High-Luminosity LHC configuration, resulting in a considerable muon rate rise, driven by both the planned luminosity increase and the enlarged magnet aperture. Nevertheless, the upgrade from emulsion films to silicon vertex detectors will preserve the efficiency of the experiment even in such a high-rate environment.
\\[1em]
\textbf{Keywords:} SND@LHC -- Muon background -- Monte Carlo simulations -- \mbox{HL-LHC}.
\end{sloppypar}
} 

\section{Introduction}
\label{sec:intro}

\begin{sloppypar}
   The Scattering and Neutrino Detector at the LHC (SND@LHC) is an experiment designed to perform measurements with high-energy neutrinos~\mbox{(100~GeV--1~TeV)} produced at the LHC in the pseudo-rapidity region \mbox{$7.2 < \eta < 8.4$}. Neutrino observations at the detector have been previously reported~\cite{SND1,SND2}. SND@LHC is situated in the TI18 service tunnel, approximately 480~m downstream of the ATLAS Interaction Point (IP1). Being shielded by roughly 100~m of rock and concrete along the ATLAS line-of-sight, long-range muons originating from proton--proton ($pp$) collisions at IP1 constitute the primary background for neutrino interaction studies. Muons may enter the detector without being vetoed and induce showers via $e^+e^-$ pair production, bremsstrahlung, or deep inelastic scattering. Alternatively, they can interact in the surrounding material, producing secondary neutral hadrons that enter the detector and potentially mimic neutrino interactions. Therefore, a comprehensive characterization of the muon flux is necessary, by means of both direct measurements and detailed Monte Carlo (MC) simulations. 
\end{sloppypar}

The aim of this work is to present the SND@LHC muon background measurements across all \mbox{Run-3} LHC configurations in comparison with the respective MC estimates, illustrating how the simulation framework helped interpret the observed variations and identify mitigation measures. Furthermore, the reported extensive benchmarking provides confidence in the predictions for the High-Luminosity LHC (\mbox{HL-LHC}) era, starting with \mbox{Run-4}.

\begin{figure*}[ht]
    \centering
    \includegraphics[width=0.85\linewidth]{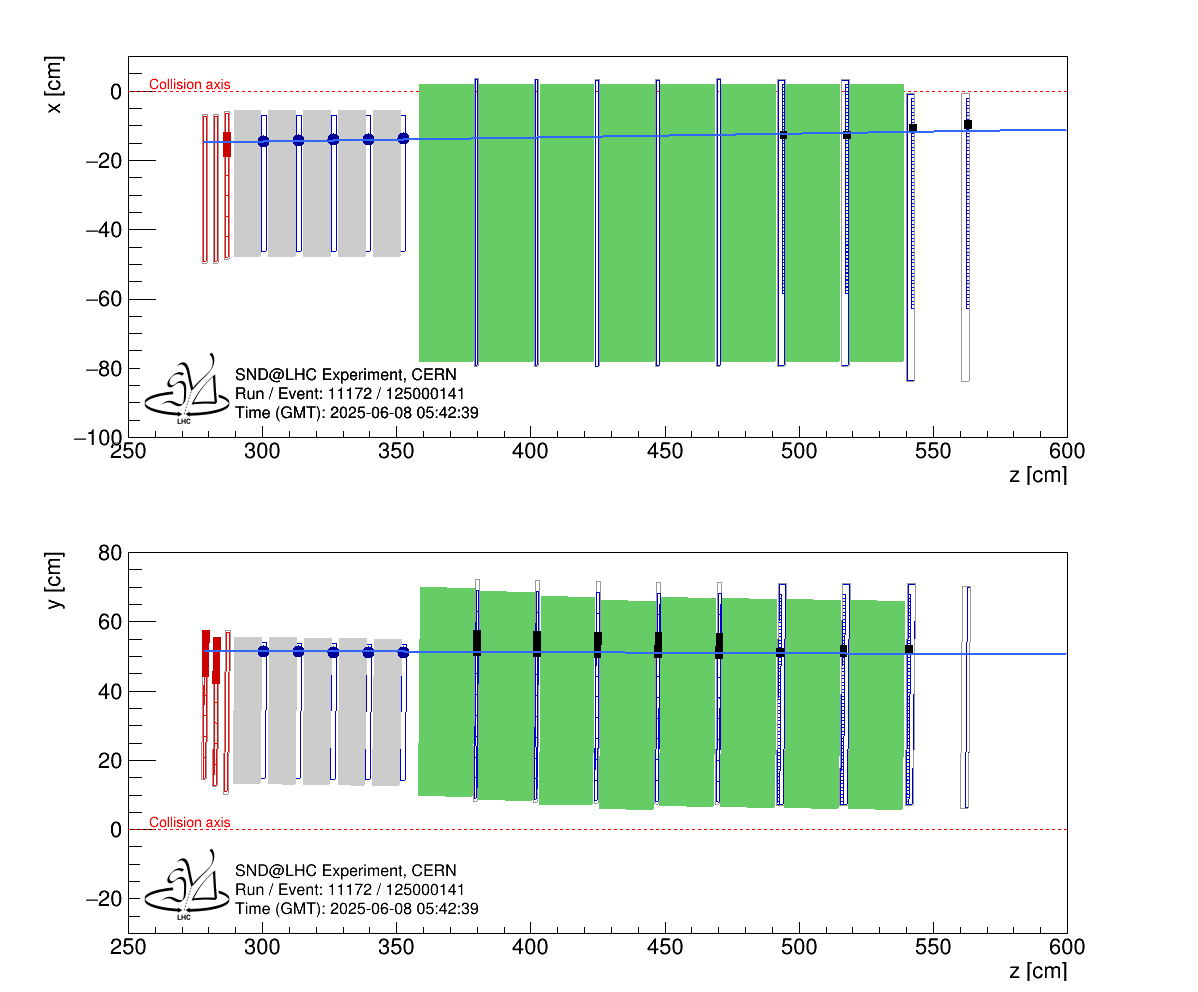}
    \caption{Representative example of a passing-through muon event. The top panel shows a top-down view of the detector and the bottom panel shows a side view.}
    \label{fig:det}
\end{figure*}

The structure of the paper is as follows. Section~\ref{sec:detector} gives a brief description of the SND@LHC detector. Next, Section~\ref{sec:simulation} introduces the relevant ingredients of the different LHC configurations during the \mbox{Run-3} operation and outlines the MC simulation chain. The evolution of the measured muon fluxes is discussed in Section~\ref{sec:ang_distrib-MC-comp} together with the interpretation provided by the MC results. Section~\ref{sec:ang_distrib} is devoted to the analysis of the muon angular distribution, which ultimately led to the conception and testing of mitigation measures. The prospects for \mbox{HL-LHC} are presented in Section~\ref{sec:HL}. Finally, in Section~\ref{sec:conclusion}, a summary and conclusions are provided.

\section{The detector}
\label{sec:detector}

SND@LHC~\cite{SND} is a compact hybrid apparatus consisting of three parts: the veto, target and muon systems (see Fig.~\ref{fig:det}). 

The veto detector is situated in front of the target region. It is made of two vertically shifted planes of seven \mbox{$42 \times 6 \times 1$~cm$^3$} scintillating bars and a third plane of vertically arranged \mbox{$46 \times 6 \times 1$~cm$^3$} bars. The latter was added in 2024 to improve the veto efficiency in identifying muons arriving from IP1~\cite{SNDveto} and the whole system was lowered into a small trench to fully cover the target.

The target section contains five walls. Each wall consists of four emulsion cloud chamber (ECC) units and is followed by a Scintillating Fibre (SciFi) tracker station. The muon system is placed downstream of the target. The electronic detectors provide the time stamp of the neutrino interaction, preselect the interaction region, and identify muons.
The sub-micrometric precision of nuclear emulsions allows the detection of short-lived particles like tau leptons. Each ECC module is a sequence of 60 emulsion films, \mbox{$19.2 \times 19.2$ cm$^2$}, interleaved with 59~1~mm-thick tungsten plates. Its weight is approximately 41.5~kg, adding up to 830~kg for the total target mass. Each SciFi station consists of one horizontal and one vertical \mbox{$39 \times 39$ cm$^2$} plane. Each
plane comprises six staggered layers of 250~$\micro$m diameter scintillating fibers.

\begin{figure*}[ht]
    \centering
    \includegraphics[width=0.95\linewidth]{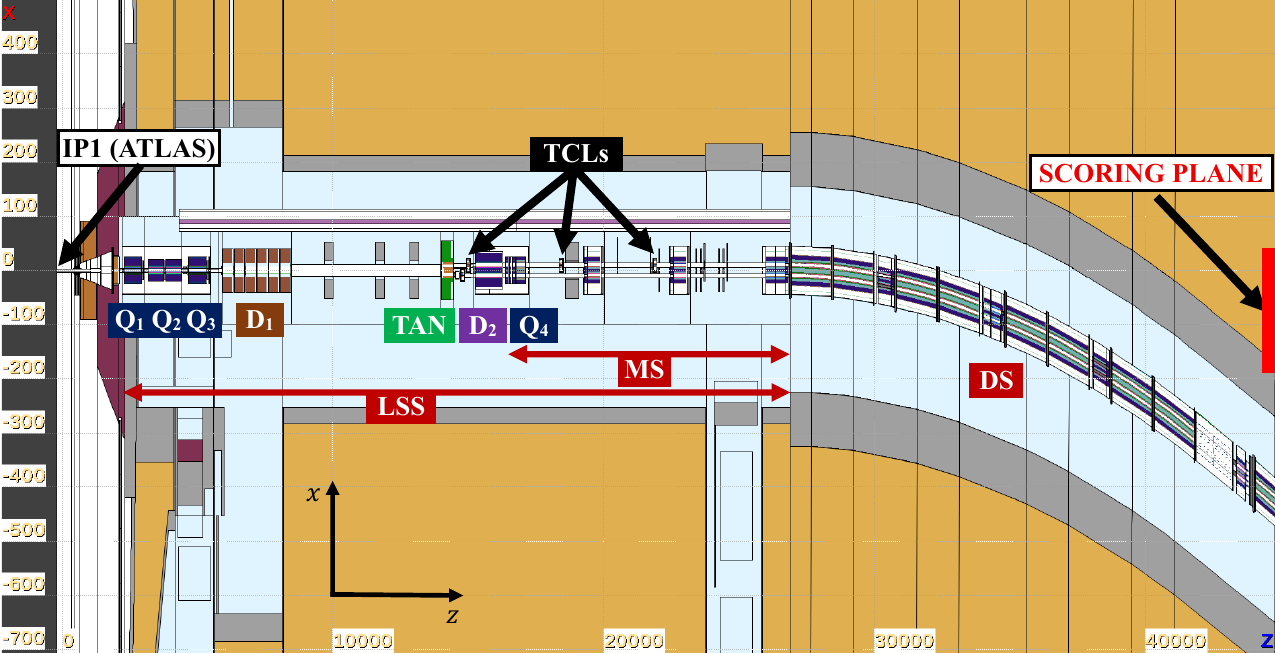}
    \caption{Top view of the FLUKA geometry model on the right side of ATLAS at beam height. 
    The transverse-to-longitudinal (x:z) scale ratio is 1:20 (values are in cm).}
    \label{fig:LHC_tunnel}
\end{figure*}

The muon system consists of two parts: upstream, the first five stations, and downstream, the last four stations. In combination with the SciFi, it acts as a coarse sampling calorimeter ($\sim$9.5~interaction lengths), providing the energy measurement of hadronic jets. Each upstream station consists of 10 stacked horizontal scintillator bars measuring \mbox{$82.5 \times 6 \times 1$~cm$^3$}. 
The downstream system comprises four planes, the first three of which each having two orthogonal layers of scintillator bars, one horizontally and one vertically arranged. A fourth plane of only vertical bars is located farthest downstream. Each bar in the downstream system has a \mbox{$1 \times 1$~cm$^2$} cross-section, with lengths of 82.5~cm for horizontal bars and 63.5~cm for vertical bars. This provides a spatial resolution of less than 1~cm in each coordinate for muon reconstruction.
The scintillating planes are interleaved with 20~cm-thick iron blocks. 
In 2025, the fourth downstream plane was moved farther downstream and two drift tube planes were installed to enhance muon tracking. The integration of data from these detectors is currently ongoing.

\section{Ruling factors and simulation framework}
\label{sec:simulation}

The muon background at SND@LHC is highly sensitive to the machine optics and beam crossing scheme implemented in Insertion Region 1 (IR1)\footnote{IR1 is the insertion region hosting the ATLAS experiment.}. Throughout \mbox{Run-3}, several distinct configurations were employed, yielding a significantly different flux and angular distribution of muons at SND@LHC, as discussed in detail in Sections~\ref{sec:ang_distrib-MC-comp}~and~\ref{sec:ang_distrib}. 

In 2022–2023, the LHC operated with nominal optics, featuring a focusing-defocusing-focusing\footnote{The focusing/defocusing effect refers to the outgoing proton beam in the horizontal plane.} (FDF) sequence for the final-focus quadrupoles (Q1, Q2, and Q3) adjacent to the ATLAS cavern (see Fig.~\ref{fig:LHC_tunnel}). Moreover, downward vertical crossing was adopted in IP1, with the momentum vector of each colliding beam tilted by an angle of $-$160~$\micro$rad in the vertical plane. In 2024, a new optics was implemented in IR1 to alter the impact distribution of the collision debris on the aforementioned quadrupoles, in this way mitigating the cumulated peak dose in the coils as a function of the produced luminosity~\cite{LHCTripletTaskForce}. Such a measure was deemed necessary to extend the machine lifetime, while exceeding the design integrated luminosity. The 2024 optics featured an inverted DFD sequence for Q1--Q3, had the Q4 quadrupole switched off, and was referred to as reverse-polarity (RP) optics. At the same time, upward vertical crossing ($+$160~$\micro$rad) was adopted in IP1. Unfortunately, the RP optics led to a substantial increase of the muon background at SND@LHC. As a consequence, in 2025, nominal optics was restored and, as an alternative measure to limit the peak dose accumulation in the coils, horizontal crossing was implemented for the first time in IP1. Although a beneficial reduction of the muon background was observed, the change of the crossing plane prevented it from returning to the 2022--2023 levels, as explained in the following sections. The optics and crossing scheme adopted in 2025 are planned to be maintained also for the 2026 operation.

The muon background also proved to be sensitive to the settings of the physics debris collimators (named TCLs) that are installed on the outgoing beam chamber along the Matching Section (MS), i.e., the accelerator segment from Q4 up to the end of the Long Straight Section (LSS) at about 270~m from IP1, as shown in Fig.~\ref{fig:LHC_tunnel}. The apertures of the TCLs jaws across the different \mbox{Run-3} years are reported in Table~\ref{tab:TCLs}.

\begin{table}[ht]
    \centering
    \caption{TCL half-gaps during \mbox{Run-3} operation.}
    \label{tab:TCLs}
    \begin{tabular}{cccc} 
    \toprule 
    \multirow{2}{*}{} & \multicolumn{3}{c}{\textbf{Half-gap (mm)}} \\ 
    \cmidrule(lr){2-4} 
                & TCL4 & TCL5 & TCL6 \\
    \midrule 
    2022--2023   & 11.5 & 14.15 & 1.835  \\
    2024        & 12.3 & 14.5  & 1.585  \\
    2025        & 12.3 & 14.5  & 1.6  \\
    \bottomrule 
    \end{tabular}
\end{table}

\begin{sloppypar}
    The MC simulation of the SND@LHC muon background relies on a two-step approach. The first step of the simulation is performed with the fully integrated MC code FLUKA~\cite{flukaweb,frontiers,batt}. The DPMJET-III event generator~\cite{dpmjet,dpmjet_phd}, embedded in FLUKA, is invoked to describe $pp$ collisions\footnote{DPMJET-III is also used in FLUKA to simulate ion--ion (\textit{e.g.}, $Pb-Pb$) nuclear inelastic collisions.} at a center-of-mass energy \mbox{$\sqrt{s} = 13.6$~TeV}. The resulting collision debris is transported through a detailed FLUKA geometry of the LHC tunnel, depicted in Fig.~\ref{fig:LHC_tunnel}, which extends from the ATLAS cavern up to a virtual interface plane, located a few tens of meters upstream of SND@LHC. Relevant characteristics of muons reaching this plane, such as kinetic energy, direction and position, are recorded, together with detailed information about their origin. The latter includes the identity of the muon parent particle and the location of the inelastic interaction that generated it. The main muon sources are the decays of light mesons (pions and kaons) produced in the primary collisions in IP1 or in subsequent secondary interactions across the LHC machine elements. To efficiently increase muon statistics, variance reduction techniques are employed in the FLUKA simulation. Specifically, the mean decay length of charged pions and charged and neutral kaons is artificially reduced to a maximum of 100~m, when physically exceeding the latter. Since muons are also produced in other (unbiased) particle decays, the consequent fluctuation of the simulated muon statistical weight is controlled by weight window biasing, keeping it within a predefined range. This prevents unphysical spikes in the resulting distributions and avoids wasting CPU time on low-importance particles. Therefore, the statistical weight of the recorded muons must be taken into account when exploiting the stored information at the interface plane.
\end{sloppypar}

The output from the FLUKA simulation serves as input for the second step of the simulation chain, which employs the Geant4 MC simulation toolkit~\cite{geant4} to model particle transport from the interface plane to and through the SND@LHC detector, with a 1~MeV kinetic energy cut-off. Only a subset of the muon population recorded by FLUKA at the scoring plane ultimately falls within the acceptance of the detector. Additionally, secondary particles may be generated by muons along their path. These are also propagated through the rock upstream of the detector, the TI18 tunnel, and the detector volume.

Initial benchmarking of the simulated muon flux for the 2025 LHC configuration against measurements at SND@LHC revealed an underestimation of $\sim$45\%. A detailed analysis (described in Section~\ref{sec:ang_distrib}) identified a significant population of negative muons originating in the Dispersion Suppressor\footnote{The region between the LSS and the LHC arc (see Fig.~\ref{fig:LHC_tunnel}).} (DS) that were not intercepted by the originally adopted interface plane. Consequently, the latter has been optimized to capture these additional muons and is set approximately 30~m upstream to the detector, with a transverse area extending over the $[-190, 40]$~cm horizontal $x$-range and the $[-90, 120]$~cm vertical $y$-range\footnote{$x=0,y=0$ is the longitudinal $z$-axis coinciding with the ATLAS detector axis (see Fig.~\ref{fig:LHC_tunnel}).}. The selected transverse area, at the indicated longitudinal position, lies outside the LHC tunnel (see Fig.~\ref{fig:LHC_tunnel}). Along their path in the rock (dry molasse, of density \mbox{$\rho \approx 2.4$~g/cm$^3$}), simulated in the second step by Geant4, minimum ionizing particles 
lose on average approximately 15~GeV. Therefore, the FLUKA simulation in the first step was performed with a conservative cut-off of 10~GeV for particle transport.
Following these optimizations, benchmarking was performed for all \mbox{Run-3} configurations, yielding a MC-estimated muon background in good agreement with the measurements at SND@LHC, as reported in the following section.

\section{Data evolution and MC interpretation}
\label{sec:ang_distrib-MC-comp}

Table~\ref{tab:muon_flux_comparison} reports the measured muon rate at SND@LHC ($\Phi_\text{exp}$) during \mbox{Run-3}, the corresponding MC estimate ($\Phi_\text{MC}$), and their relative difference, evaluated as
\begin{equation}
    \delta = \frac{\Phi_\text{MC} - \Phi_\text{exp}}{\Phi_\text{exp}}.
\end{equation}

\begin{table}[ht]
    \centering
    \caption{SND@LHC muon background during the \mbox{Run-3} years: measured and simulated rates at \mbox{$\mathcal{L} = 2 \times 10^{34}$~cm$^{-2}$~s$^{-1}$}, with respective statistical uncertainties, and percentage differences.}
    \label{tab:muon_flux_comparison}
    \begin{tabular}{lccc} 
        \toprule
        & \textbf{2022--2023} & \textbf{2024} & \textbf{2025} \\ 
        \midrule
        $\Phi_\text{exp}$ (Hz) & 557 $\pm$ 1  & 1154 $\pm$ 2  & 799 $\pm$ 1 \\ 
        $\Phi_\text{MC}$ (Hz)  & 500 $\pm$ 17 & 1008 $\pm$ 35 & 907 $\pm$ 40 \\ 
        $\delta$               & -10\%        & -12.5\%       & +13.5\% \\ 
        \bottomrule
    \end{tabular}
\end{table}

\begin{figure}
    \centering
    \includegraphics[width=0.825\columnwidth]{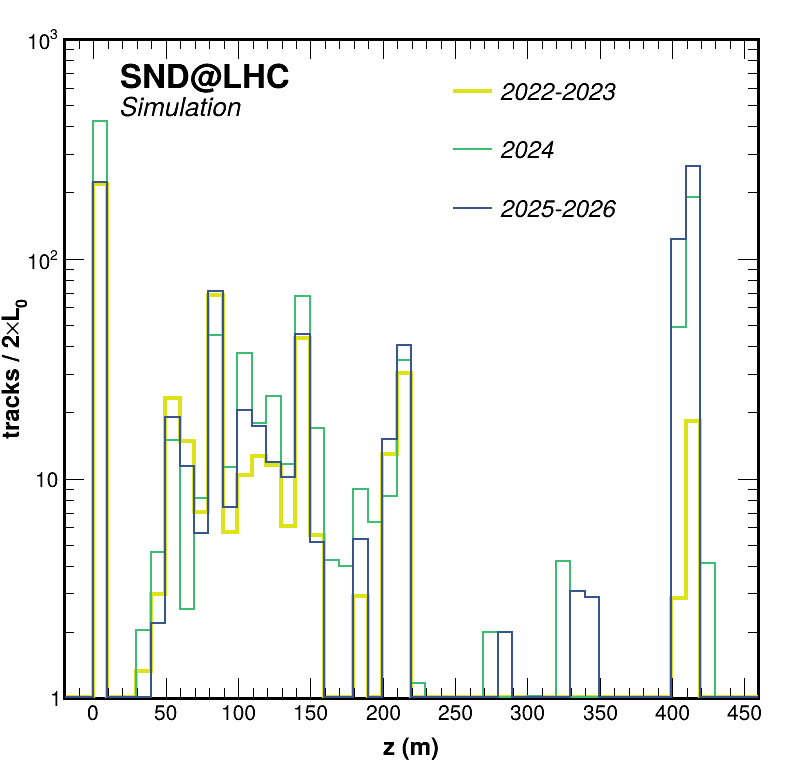}
    \caption{Distribution of the origin of the muons reaching SND@LHC in the simulation for the three indicated \mbox{Run-3} configurations: 2022--2023 in yellow, 2024 in green, and 2025--2026 in blue. The $z$-coordinate refers to the location of the last nuclear reaction preceding the muon generation. See Fig.~\ref{fig:LHC_tunnel} for the correspondence with the machine elements. \mbox{$L_0 = 10^{34}$~cm$^{-2}$~s$^{-1}$} is the LHC design instantaneous luminosity, which was routinely doubled during \mbox{Run-3}.}
    \label{fig:z_origin}
\end{figure}

\begin{figure}
    \centering
    \includegraphics[width=0.85\columnwidth]{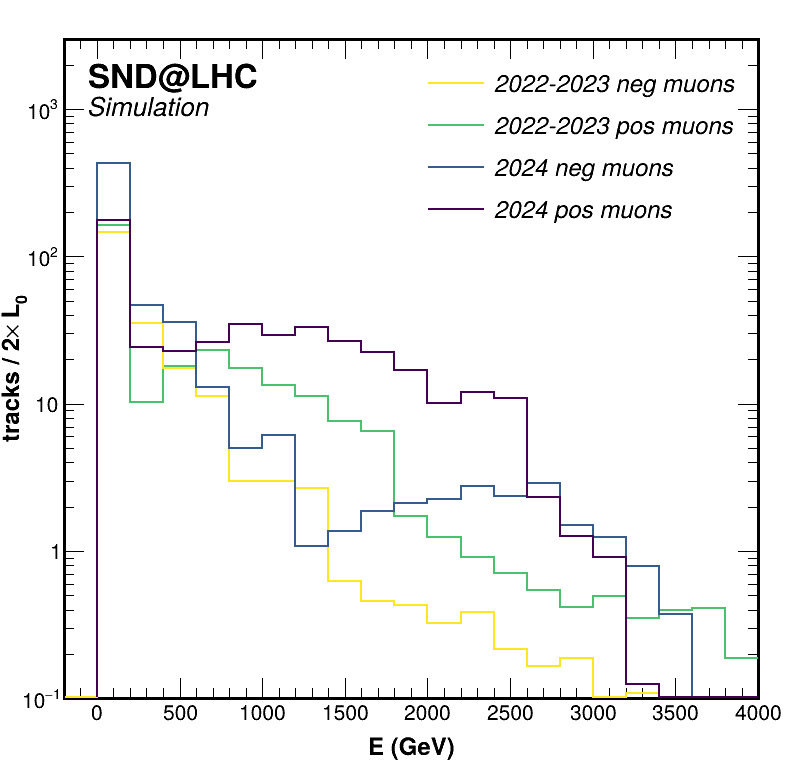}
    \caption{Simulated negative and positive muon spectra at SND@LHC for the 2022--2023 and 2024 configurations. \mbox{$L_0 = 10^{34}$~cm$^{-2}$~s$^{-1}$} is the LHC design instantaneous luminosity, which was routinely doubled during \mbox{Run-3}.}
    \label{fig:mu_spectrum}
\end{figure}

The experimental muon rate is determined with the SciFi tracker, whose acceptance matches the ECC one. Track reconstruction is performed by means of the Hough Transform pattern recognition approach~\cite{hough}. Track candidates are fitted using the Kalman filter~\cite{kalman} implemented in the GENFIT package~\cite{genfit}. The 3D trajectories are constructed by combining 2D projections in the $xz$ and $yz$ planes. The muon rates reported in this work include tracks over the entire $39 \times 39 \text{ cm}^2$ target region, corresponding to the full ECC acceptance. This definition facilitates the planning of emulsion exposure and replacement during data-taking. The collaboration has previously reported muon rates for selected detector areas with refined tracking efficiency~\cite{SND2024,SND2026}.

The MC muon rate at SND@LHC is estimated as follows:
\begin{equation}
    \Phi_\text{MC} = \cfrac{1}{N_{pp}} \left(\sum_{i=1}^{N_\text{tracks}} w_i\right) \times \sigma_\text{inel} \times \mathcal{L},
\end{equation}
where $N_{pp}$ is the number of $pp$ collisions simulated in the first step by FLUKA (of the order of 50--100~millions), $N_\text{tracks}$ is the total number of muons reconstructed within the active volume of the detector, and the $w_i$ are their respective statistical weights. Furthermore, \mbox{$\sigma_\text{inel} \approx 80\text{ mb}$} denotes the inelastic proton--proton cross-section at \mbox{$\sqrt{s} = 13.6\text{ TeV}$} (including diffractive events), while \mbox{$\mathcal{L} = 2 \times 10^{34}$~cm$^{-2}$~s$^{-1}$} represents the reference instantaneous luminosity.

As anticipated in the previous section and quantified in Table~\ref{tab:muon_flux_comparison}, the muon background measured at SND@LHC exhibited significant variations across \mbox{Run-3} depending on the machine optics and beam crossing scheme adopted in IR1. The MC predictions reproduce well the trend and are in satisfactory agreement with the measurements, in spite of the complexity of the simulation chain.
As an additional insight, the latter allows for the interpretation of the aforementioned variations, thanks to the information on the origin of the muons reaching SND@LHC. Figure~\ref{fig:z_origin} shows the distribution of the longitudinal position where the nuclear reaction generating the parent particle of the detected muon took place in the simulation.
In 2024, the muon flux increased by approximately a factor of two compared to the previous years. This rise was driven by muons originating from decaying particles directly produced by the primary collisions inside ATLAS, as indicated by the respective increase of the main peak on the left of the plot. The rise of this component was a consequence of the inverted DFD sequence applied to the Q1–Q3 quadrupoles with the RP optics. Figure~\ref{fig:mu_spectrum} highlights that the latter favored the arrival of energetic ($>1$~TeV) positive muons at the detector. A concurrent augmentation in low-energy negative muons can also be observed, but it has a different origin, as explained in the following. In 2025, the re-establishment of nominal optics reduced the background with respect to 2024, but did not restore the 2022--2023 levels. As shown on the right of Fig.~\ref{fig:z_origin}, the difference was driven by the negative muon population originating from the DS, where the adoption of horizontal crossing in IP1 maximized the loss of diffractive protons close to beam energy, already boosted by the RP optics in 2024 (with vertical crossing). These protons impact on the half-cell 11 magnets, at 400~m from IP1 (where the single-turn dispersion peaks), generating muons that, if negatively charged, are bent by the dipole field towards the detector. This component can be clearly identified in the angular distribution of muon tracks at SND@LHC, as elucidated in the following section.

\section{Muon angular distribution}
\label{sec:ang_distrib}

\begin{figure}
    \centering
    \includegraphics[width=0.85\columnwidth]{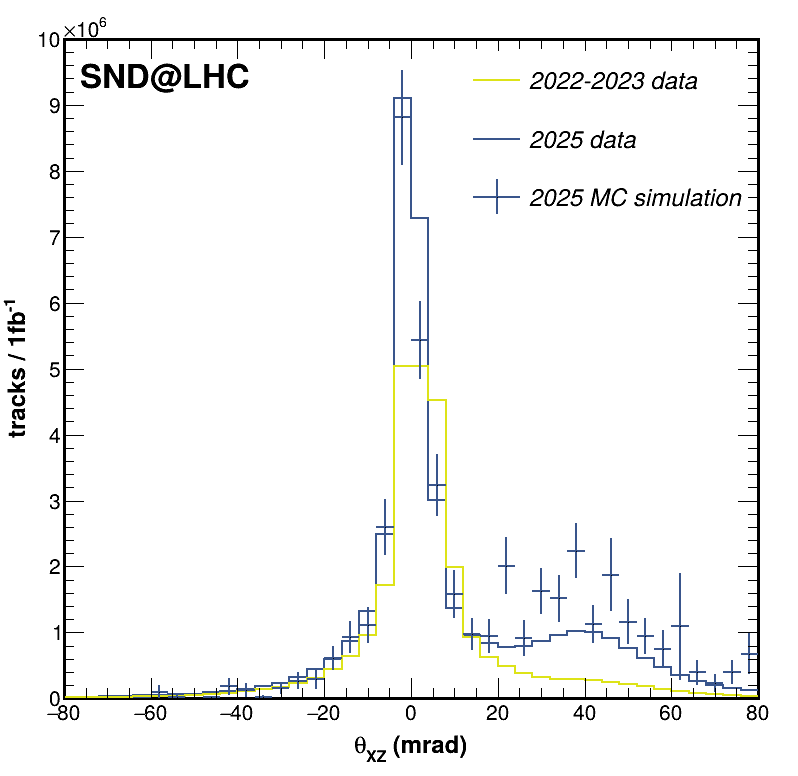}
    \caption{Measured horizontal angular distribution of muons reaching SND@LHC in 2022--2023 (yellow) and 2025 (blue), compared to the MC prediction for 2025 (points).}
    \label{fig:ang2325.png}
\end{figure}

\begin{figure}
    \centering
    \includegraphics[width=0.85\columnwidth]{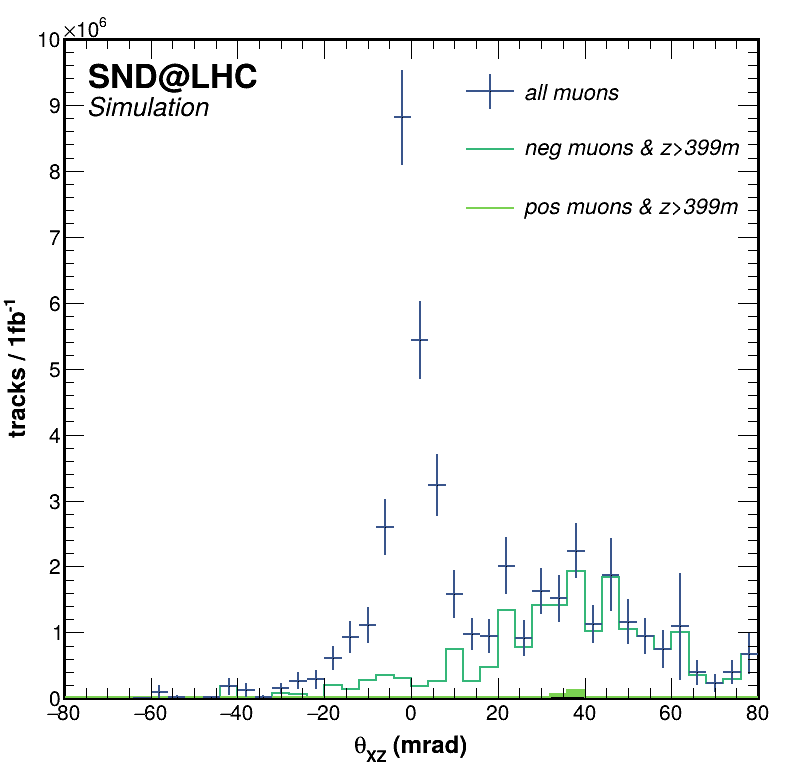}
    \caption{MC prediction for the horizontal angular distribution of muons in 2025 (points) isolating the contribution of negative (dark green) and positive (light green) muons originating in half-cell 11.}
    \label{fig:ang25MC.png}
\end{figure}

Figure~\ref{fig:ang2325.png} shows the horizontal angular distributions of muons measured by SND@LHC in 2022--2023 and 2025, respectively (2024 is omitted from the comparison in view of the different optics configuration). The 2025 curve displays a pronounced bump at large positive angles, in the \mbox{20--60~mrad} range, corresponding to muons directed outside the LHC ring. The 2025 simulation, while overestimating this component, clearly indicates that it is due to negative muons deriving from nuclear reactions taking place mostly in the second main dipole of half-cell 11 (see Fig.~\ref{fig:ang25MC.png}), which explains their large angle at the detector. These muons fell outside the acceptance of a first interface plane designed in the simulation chain, leading to the significant underestimation recalled in Section~\ref{sec:simulation} and to the optimized redefinition of the interface plane.

\begin{figure}
    \centering
    \includegraphics[width=\columnwidth]{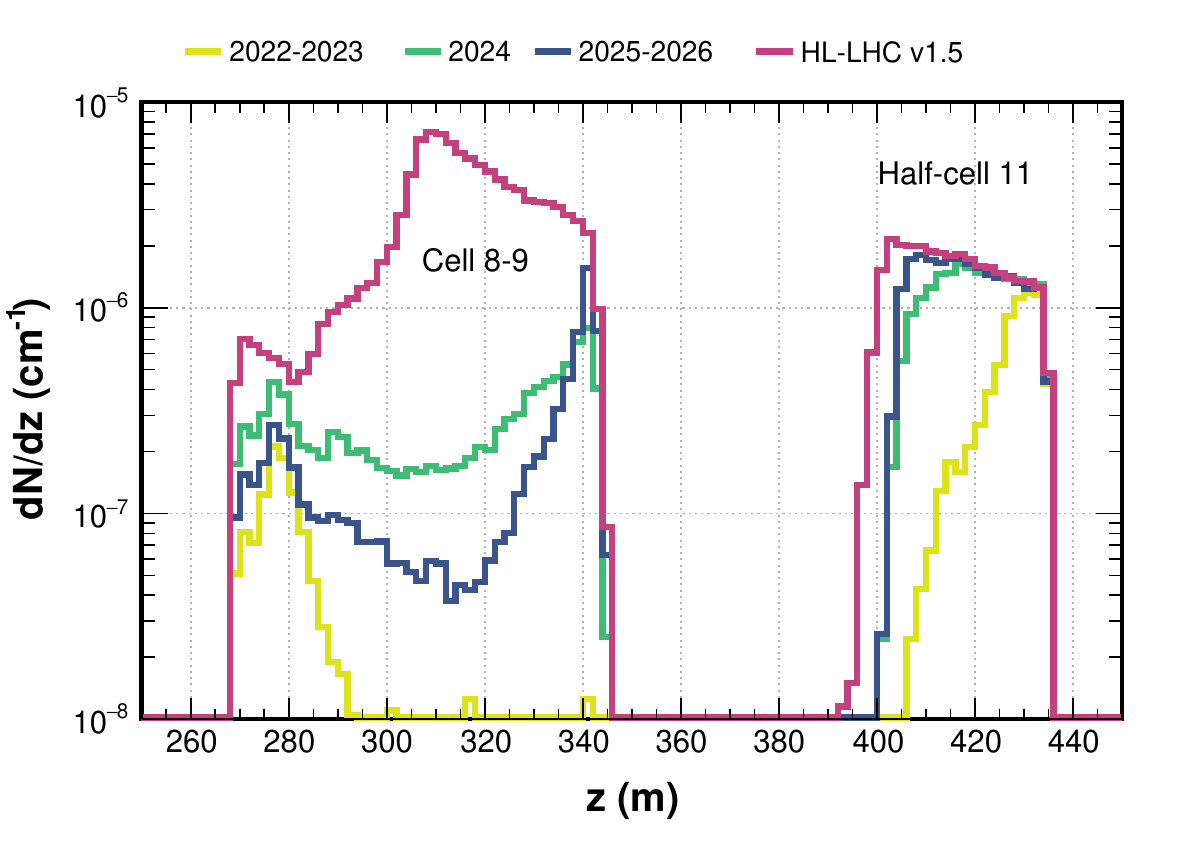}
    \caption{Simulated diffractive proton losses in the DS for the different LHC \mbox{Run-3} configurations and \mbox{HL-LHC} \mbox{Run-4}, normalized to one $pp$ collision in IP1.}
    \label{fig:DS_losses}
\end{figure}

\begin{figure}
    \centering
    \includegraphics[width=\columnwidth]{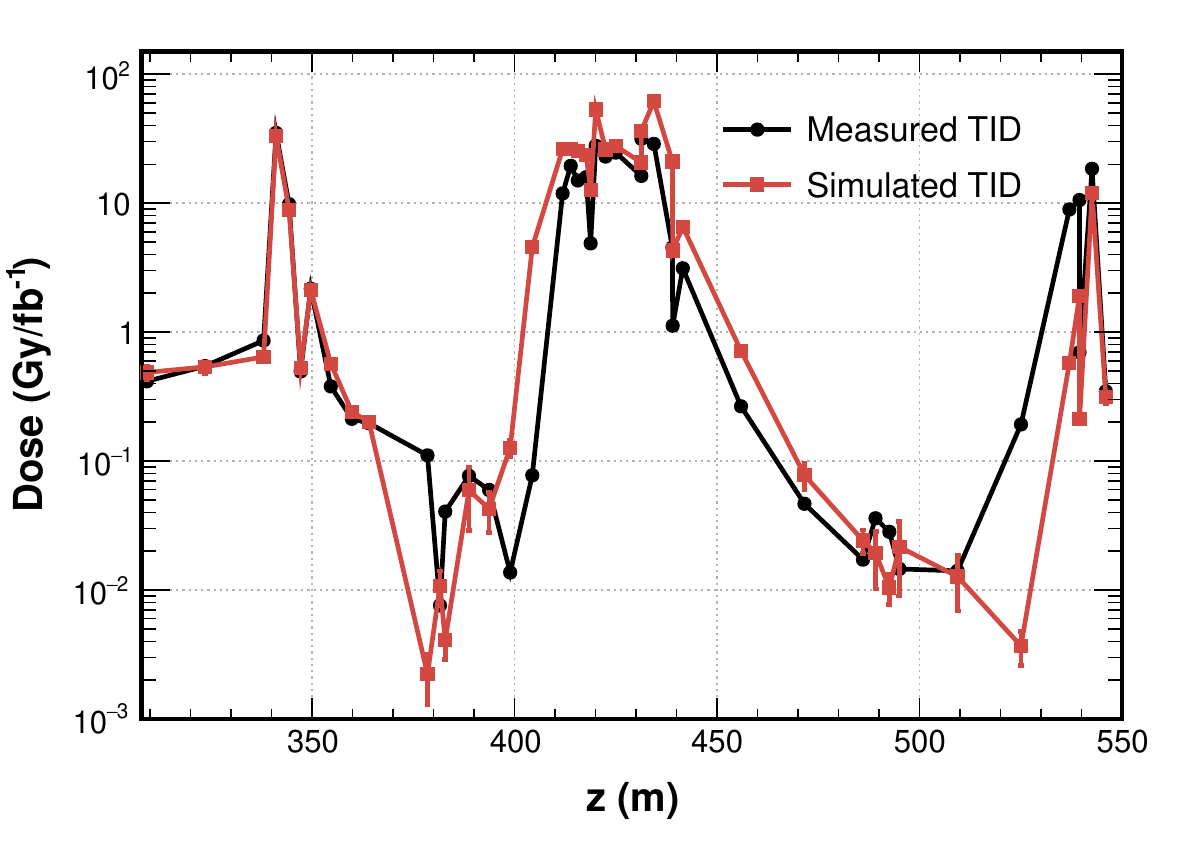}
    \caption{Measured (black) and simulated (red) pattern of the BLM signals, expressed as Total Ionizing Dose (TID), in the DS for the 2025 LHC run.}
    \label{fig:BLM_2025}
\end{figure}

As noted earlier, this muon source close to the detector is due to beam protons diffracted in IP1 that exit the vacuum chamber 400~m downstream, where their separation from the beam orbit is maximized. Figure~\ref{fig:DS_losses} illustrates the evolution of two proton loss clusters in cell 8-9 and half-cell 11 through the different \mbox{Run-3} configurations and in \mbox{Run-4}, when the IR1 layout will be transformed by the \mbox{HL-LHC} upgrade, as discussed in the next section. In particular, the marked increase from the 2022--2023 distribution to the 2025--2026 one is apparent, as a result of the transition from vertical to horizontal crossing. A similar increase from 2022--2023 can be appreciated for 2024, as an RP optics effect. The reliability of the MC predictions of Fig.~\ref{fig:DS_losses}
is proven by the absolute comparison between simulated and measured Beam Loss Monitor (BLM) doses, reported in Fig.~\ref{fig:BLM_2025} for the 2025 case. Despite a noteworthy global agreement, which is excellent in half-cell 9, a local overestimation of up to a factor of 2 is observed at the half-cell 11 maximum. This can explain the already underlined excess in the positive-angle tail of the MC muon angular distribution for 2025, further quantified in Table~\ref{tab:percentage}.

\begin{table}[ht]
    \centering
    \caption{Percentage of muons in various $\theta_{xz}$ angular ranges for the 2025 LHC configuration.}
    \label{tab:percentage}
    \renewcommand{\arraystretch}{1.25} 
    \begin{tabular}{lccc}
    \toprule
    \multirow{3}{*}{\textbf{2025 (nominal)}} & \multicolumn{3}{c}{$\theta_{xz}$~(mrad)} \\ 
    \cmidrule(lr){2-4}
    & $< -10$ & $[-10, 10]$ & $> 10$ \\ 
    \cmidrule(lr){2-4}
    & \multicolumn{3}{c}{\textbf{\% of muons}} \\ 
    \midrule
    \textbf{Measurement} & 11 & 59 & 30  \\ 
    \textbf{MC}          & 7  & 48 & 45  \\ 
    \bottomrule
    \end{tabular}
\end{table}

Nonetheless, the insight provided by the simulation chain triggered the conception of mitigation measures, which were successfully tested on two occasions. Considering that the DS contribution to the muon background proved to be dominated by proton losses in half-cell 11, with the cell 8-9 cluster playing only a minor role, the idea was to alter the beam optics so as to displace the half-cell 11 cluster either downstream or upstream. Therefore, respective orbit bumps were designed~\cite{BE-ABP} and implemented. As a result, a 9~mm orbit bump, moving the half-cell 11 cluster into half-cells 13 and 15, yielded a measured 20\% reduction of the muon background at SND@LHC. An alternative 8~mm orbit bump, moving the cluster into cell 8-9, reduced the measured background by 15\%.
These reductions were overestimated by the simulation because of the MC overestimation of the half-cell 11 contribution previously discussed. Eventually, no orbit bump was retained for the remaining duration of \mbox{Run-3}, due to the challenges of their regular adoption and especially the limited target luminosity for 2026. 


\section{The \mbox{HL-LHC} picture}
\label{sec:HL}

The primary objective of the \mbox{HL-LHC} upgrade is to raise the LHC integrated luminosity up to a target of roughly $3000$~fb$^{-1}$ over a decade~\cite{hllhc_tdr}, after resuming operation in 2030, through the increase of the instantaneous luminosity in ATLAS and CMS up to a nominal value of \mbox{$5 \times 10^{34}\text{ cm}^{-2}\text{s}^{-1}$}. For SND@LHC, this implies a significant increase of both signal and background, with the evolution of their ratio to be accurately characterized.

\begin{figure*}
    \centering
    \includegraphics[width=0.85\linewidth]{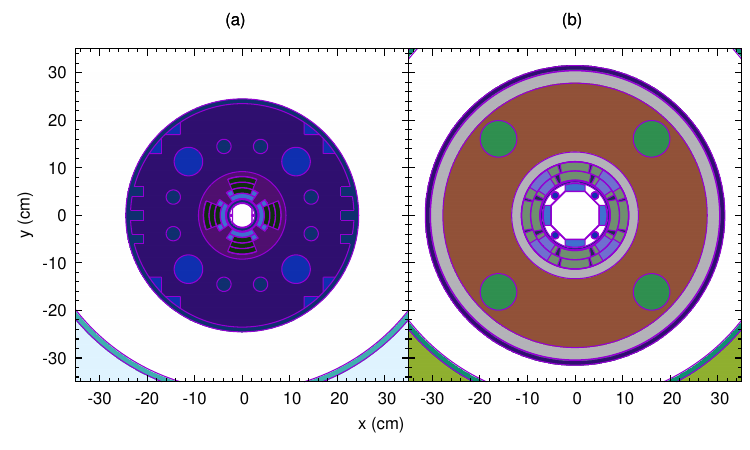}
    \caption{Transverse cross-section of the FLUKA geometry model of the Q1 final-focus quadrupole magnet in (a) LHC and (b) \mbox{HL-LHC}.}
    \label{fig:Q1}
\end{figure*}

As anticipated in Fig.~\ref{fig:DS_losses}, diffractive proton losses in the IR1 DS are expected to remain sizeable during the \mbox{HL-LHC} era, given the horizontal baseline crossing plane in ATLAS, and exhibit a considerable augmentation in cell 8-9, for the assumed TCL6 half-gap of 3~mm. The application of the same two-step simulation framework to the \mbox{HL-LHC} geometry model in FLUKA indicates that the cell 8-9 cluster, despite its prominence, comprises less than 10\% of the muon background at SND@LHC. On the other hand, proton losses in half-cell 11 continue to play a substantial role, accounting for 35\% of the MC muon flux at the detector (to be compared to the 2025 percentages in Table~\ref{tab:percentage}). Therefore, the use of orbit bumps, such as those successfully tested in 2025, to suitably displace the DS losses, remains of interest as a mitigation strategy for \mbox{Run-4}. 

\begin{sloppypar}
    When normalizing the \mbox{HL-LHC} background prediction to the LHC instantaneous luminosity \mbox{$\mathcal{L} = 2 \times 10^{34}$~cm$^{-2}$~s$^{-1}$} (rather than the \mbox{HL-LHC} target, which is a factor of 2.5 higher), the resulting muon rate is 1338~Hz (in reality expected to rise to 3345~Hz). This exceeds the highest muon rate to date that was experienced in 2024 (see Table~\ref{tab:muon_flux_comparison}), although the \mbox{HL-LHC} configuration features the nominal optics (FDF) and not the RP optics (DFD). On the other hand, apart from the increase of the half-crossing angle magnitude from 160 to 250~$\micro$rad, major layout differences distinguish \mbox{HL-LHC} from LHC in IR1. In particular, the coil aperture of the final-focus quadrupoles gets enlarged from 70 to 150~mm in diameter, as shown in Fig.~\ref{fig:Q1}, and their total length increases by over 10~m. Furthermore, the present normal-conducting separation dipole D1, made of 6 identical magnets with a vertical aperture of 63~mm, will be replaced by a single superconducting magnet having the same 150~mm coil aperture as the new quadrupoles. Simulations with the aperture of the Q1–D1 string artificially reduced to a 70~mm diameter yield a decrease of the muon background by a factor of 2, emphasizing the role of the mechanical machine aperture.

\end{sloppypar}

Unfortunately, mitigation measures for muons originating along the LSS are far from straightforward. However, the predicted background rate, albeit considerably higher, is expected to be sustainable. In fact, the primary constraint during \mbox{Run-3} was related to the use of emulsion films. Their planned replacement with a silicon vertex detector~\cite{snd_hl_tdr} should enable efficient \mbox{HL-LHC} operation even with such a high muon background. It has to be added that the \mbox{HL-LHC} alternative scenario featuring vertical crossing in ATLAS (either upwards or downwards) would decrease the muon background at SND@LHC by approximately 20\% compared to the baseline horizontal crossing. While this scenario remains possible, its implementation requires the physical swap of crab cavities between IR1 and IR5, which is a demanding intervention. 

\section{Conclusion}
\label{sec:conclusion}

The systematic evaluation of the muon flux at SND@LHC through both measurements and Monte Carlo simulations is essential, as it constitutes the primary background for neutrino interaction searches and directly impacts the replacement frequency of the emulsion target system.

The study presented in this work correlates differences in the flux and angular distribution of muons at SND@LHC with the different LHC configurations implemented in Insertion Region 1 during \mbox{Run-3}. While the 2024 reverse-polarity optics resulted in a twofold increase of the muon background, the 2025 transition to horizontal crossing with nominal optics granted only a partial reduction. In fact, the muon rate remained above the 2022–2023 levels due to larger diffractive proton losses in the Dispersion Suppressor region, as unveiled by simulation. As a mitigation of this effect, orbit bumps displacing the proton losses from half-cell 11 reduced the measured background by 15–20\%.

An initial discrepancy between the Monte Carlo estimate and the 2025 measurement revealed the role of the aforementioned losses as a source of large-angle (\mbox{$\theta_{xz} > 10$~mrad}) negative muons reaching SND@LHC. By moving the interface plane between the two steps of the simulation chain closer to the detector and suitably expanding its transverse dimensions, an agreement within 10–15\% between the measured and simulated muon fluxes was achieved for all \mbox{Run-3} configurations.

With respect to 2025, the prediction for the \mbox{HL-LHC} era features a fourfold increase in the muon rate, which is expected to exceed 3~kHz at the new target instantaneous luminosity. This rise is driven by both the luminosity upgrade and the larger aperture of the \mbox{HL-LHC} final-focus quadrupoles and separation dipole. Nonetheless, the planned transition from emulsion films to silicon vertex detectors will allow the SND@LHC experiment to withstand such a high background.

\section*{Acknowledgements}
We thank V. Rodin for the setup of the different LHC configurations in FLUKA, M. Sabat\'e Gilarte and D. Prelipcean for the \mbox{HL-LHC} model, and D. Soderstrom for making readily available the BLM data. We express our gratitude to R. Bruce, A. Donadon Servelle, and B. Lindstrom for the conception of the beam orbit bumps and to the LHC operation team for the implementation of the latter.

We acknowledge the support for the construction and operation of the SND@LHC detector provided by the following funding agencies: CERN; the Bulgarian Ministry of Education and Science within the National
Roadmap for Research Infrastructures 2020–2027 (object CERN) and for the support of this study under the National Program ``Young Researchers and Posdoctoral Students - 2''; ANID FONDECYT grants No. 3230806, No. 1240066, 1240216 and ANID  - Millenium Science Initiative Program -  $\rm{ICN}2019\_044$ (Chile); the Deutsche Forschungsgemeinschaft (DFG, ID 496466340); the Italian National Institute for Nuclear Physics (INFN); JSPS, MEXT, the~Global COE program of Nagoya University, the~Promotion and Mutual Aid Corporation for Private Schools of Japan for Japan; the National Research Foundation of Korea with grant numbers 
2021R1A2C2011003,\\ 2020R1A2C1099546, 2021R1F1A1061717, and \\
2022R1A2C100505; Fundação para a Ciência e a Tecnologia, FCT grant numbers  CEECIND/01334/2018, CEECINST/00032/2021 and PRT/BD/153351/2021\\
(Portugal), CERN/FIS-INS/0028/2021; the Swiss National Science Foundation (SNSF); TENMAK for Turkey (Grant No. 2022TENMAK(CERN) A5.H3.F2-1). \\
J.C.~Helo~Herrera and O.~J.~Soto~Sandoval acknowledge support from ANID FONDECYT grants No.1241685 and 1241803. M.~Climescu, H.~Lacker and R.~Wanke are funded by the Deutsche Forschungsgemeinschaft (DFG, German Research Foundation), Project 496466340. This research was financially supported by the Italian Ministry of University and Research within the Prin 2022 program.

We express our gratitude to our colleagues in the CERN accelerator departments for the excellent performance of the LHC. We thank the technical and administrative staff at CERN and at other SND@LHC institutes for their contributions to the success of the SND@LHC efforts. We thank Luis Lopes, Jakob Paul Schmidt and Maik Daniels for their help during the~construction.

\bibliographystyle{spphys}
\bibliography{ref}

\clearpage
\onecolumn
\onecolumn
\begin{center}
\textbf{The SND@LHC Collaboration}
\vspace{0.25cm}
\break
\author{D.~Abbaneo$^{9}$\orcidlink{0000-0001-9416-1742}},
\author{S.~Ahmad$^{42}$\orcidlink{0000-0001-8236-6134}},
\author{R.~Albanese$^{1,2}$\orcidlink{0000-0003-4586-8068}},
\author{A.~Alexandrov$^{1}$\orcidlink{0000-0002-1813-1485}},
\author{F.~Alicante$^{1,2}$\orcidlink{0009-0003-3240-830X}},
\author{F.~Aloschi$^{1,2}$\orcidlink{0000-0002-2501-7525}},
\author{K.~Androsov$^{6}$\orcidlink{0000-0003-2694-6542}},
\author{A.~Anokhina$^{3}$\orcidlink{0000-0002-4654-4535}},
\author{L.G.~Arellano$^{1,2}$\orcidlink{0000-0002-1093-1824}},
\author{C.~Asawatangtrakuldee$^{38}$\orcidlink{0000-0003-2234-7219}},
% ...
\end{center}

\begin{flushleft}
\begin{footnotesize}
$^{1}$Sezione INFN di Napoli, Napoli, 80126, Italy\linebreak
$^{2}$Universit\`{a} di Napoli ``Federico II'', Napoli, 80126, Italy\linebreak
$^{3}$Affiliated with an institute formerly covered by a cooperation agreement with CERN\linebreak
% ...
$^{6}$Institute of Physics, EPFL, Lausanne, 1015, Switzerland\linebreak
% ...
$^{9}$European Organization for Nuclear Research (CERN), Geneva, 1211, Switzerland\linebreak
% ...
$^{38}$Department of Physics, Faculty of Science, Chulalongkorn University, Bangkok, 10330, Thailand\linebreak
% ...
$^{42}$Affiliated with Pakistan Institute of Nuclear Science and Technology (PINSTECH), Nilore, 45650, Islamabad, Pakistan\linebreak
% ...
\end{footnotesize}
\end{flushleft}

\end{document}